**Magnetospheric Multiscale Observations of Electron Vortex Magnetic Hole in the Magnetosheath Turbulent Plasma**

S. Y. Huang[1,2], F. Sahraoui[2], Z. G. Yuan[1], J. S. He[3], J. S. Zhao[4], O. Le Contel[2], X. H. Deng[5], M. Zhou[6], H. S. Fu[7], Y. Pang[5], Q. Q. Shi[8], B. Lavraud[9,10], J. Yang[7], D. D. Wang[1], X. D. Yu[1], C. J. Pollock[11], B. L. Giles[11], R. B. Torbert[12], C. T. Russell[13], K. A. Goodrich[14], D. J. Gershman[11], T. E. Moore[11], R. E. Ergun[14], Y. V. Khotyaintsev[15], P.-A. Lindqvist[16], R. J. Strangeway[13], W. Magnes[17], K. Bromund[18], H. Leinweber[13], F. Plaschke[17], B. J. Anderson[18], and J. L. Burch[19]

[1] School of Electronic Information, Wuhan University, Wuhan, China

[2] Laboratoire de Physique des Plasmas, CNRS-Ecole Polytechnique-UPMC, Palaiseau, France

[3] School of Earth and Space Sciences, Peking University, Beijing, China

[4] Key Laboratory of Planetary Sciences, Purple Mountain Observatory, Chinese Academy of Sciences, Nanjing, China

[5] Institute of Space Science and Technology, Nanchang University, Nanchang, China

[6] Department of Physics and Astronomy, University of California, Los Angeles, CA, USA

[7] School of Space and Environment, Beihang University, Beijing, China

[8] Shandong Provincial Key Laboratory of Optical Astronomy and Solar-Terrestrial Environment, Institute of Space Sciences, Shandong University, Weihai, China

[9] Institut de Recherche and Astrophysique et Planétologie, Université de Toulouse (UPS), Toulouse, France

[10] Centre National de la Recherche Scientifique, UMR 5277, Toulouse, France

[11] NASA, Goddard Space Flight Center, Greenbelt, Maryland, USA

[12] University of New Hampshire, Durham, New Hampshire, USA

[13] Department of Earth, Planetary and Space Sciences, University of California, Los Angeles, CA, USA

[14] University of Colorado, Boulder, Colorado, USA

[15] Swedish Institute of Space Physics, Uppsala, Sweden

[16] Royal Institute of Technology, Stockholm, Sweden

[17] Space Research Institute, Austrian Academy of Sciences, Graz, Austria

[18] Johns Hopkins University Applied Physics Laboratory, Laurel, MD, USA

[19] Southwest Research Institute, San Antonio TX, USA

**Abstract**

We report the observations of an electron vortex magnetic hole corresponding to a new type of coherent structures in the magnetosheath turbulent plasma using the Magnetospheric Multiscale (MMS) mission data. The magnetic hole is characterized by a magnetic depression, a density peak, a total electron temperature increase (with a parallel temperature decrease but a perpendicular temperature increase), and strong currents carried by the electrons. The current has a dip in the center of the magnetic hole and a peak in the outer region of the magnetic hole. The estimated size of the magnetic hole is about 0.23 $\rho_i$ (~ 30 $\rho_e$) in the circular cross-section perpendicular to its axis, where $\rho_i$ and $\rho_e$ are respectively the proton and electron gyroradius. There are no clear enhancement seen in high energy electron fluxes, but an enhancement in the perpendicular electron fluxes at ~ 90° pitch angles inside the magnetic hole is seen, implying that the electron are trapped within it. The variations of the electron velocity components $V_{em}$ and $V_{en}$ suggest that an electron vortex is formed by trapping electrons inside the magnetic hole in the circular cross-section (in the M-N plane). These observations demonstrate the existence of a new type of coherent structures behaving as an electron vortex magnetic hole in turbulent space plasmas as predicted by recent kinetic simulations.

## 1. Introduction

Turbulence is an ubiquitous feature of various space and astrophysical plasmas that include the solar wind, the planetary magnetospheres, the interstellar medium and accretion flows (e.g., Tu and Marsch 1995; Bruno and Carbone 2005; Sahraoui et al. 2006, 2009, 2010, 2013; Scheckochihin et al. 2009; Huang et al. 2012, 2014). Nonlinear energy cascade in magnetized turbulent plasmas leads of the formation of different coherent structures such as mirror modes (Sahraoui et al., 2004; 2006), current sheets and other discontinuities (e.g., Retino et al. 2007; Chasapis et al. 2015, Osman et al. 2012; Wang et al. 2013; Yang et al. 2015), and magnetic islands/flux ropes (e.g., Karimabadi et al. 2014; Huang et al. 2016a, 2016b). These coherent structures are thought to play an important role in dissipating energy and transporting particles in turbulent plasmas (e.g., Sundkvist et al. 2007; Servidio et al. 2014; Zhang et al. 2015).

Recent kinetic simulations revealed the existence of a new type of nonlinear coherent structure in magnetized plasma turbulence: electron vortex magnetic holes (Haynes et al. 2015; Roytershteyn et al. 2015). The electron vortex magnetic hole can grow self-consistently in the decaying turbulence, and is not the consequence of mirror mode or field swelling instability growth and saturation as shown in two-dimensional and three-dimensional particle-in-cell (PIC) simulations (Haynes et al. 2015; Roytershteyn et al. 2015). Those simulations showed the following basic properties of the electron vortex magnetic hole: size of the order of the electron scale, circular cross section, trapping of electrons to form the electron vortex, and an electron perpendicular temperature larger than the parallel one.

Magnetic holes with significant magnetic field depression (or reduction) have been reported in the solar wind (e.g., Turner et al. 1977; Zhang et al. 2008), in the magnetosheath (e.g., Tsurutani et al. 2011; Yao et al., 2017) and in the magnetotail plasma sheet (e.g., Ge et al. 2011). Recently, sub-proton scale magnetic holes were detected in the magnetotail using the data of THEMIS, Cluster and MMS (Magnetospheric Multiscale) missions (e.g., Sun et al. 2012; Sundberg et al. 2015; Gershman et al., 2016; Goodrich et al. 2016a, 2016b). However, before the launch of MMS the low time resolution of particle data limited the characterization of magnetic holes and did not allow us to accurately identify electron vortex magnetic holes. In this Letter, we provide evidence of the existence of the electron vortex magnetic hole in the turbulent magnetosheath and perform an in-depth analysis of its dynamics thanks to the unprecedented high time resolution data of the MMS mission [*Burch et al.*, 2015].

## 2. MMS Observations of Electron Vortex Magnetic Hole

The MMS spacecraft, launched in March 2015, consists of four identical spacecraft that provide unprecedented high time resolution of the plasma data. Here we analyze burst mode data from MMS measured in magnetosheath turbulent plasma. We use the magnetic field data from the Fluxgate Magnetometer (FGM) sampled at 128 Hz (*Russell et al.* 2014), and the 3D particle data (distribution functions and plasma moments) coming from the Fast Plasma Instrument (FPI) sampled each 30 ms for electrons and 150 ms for ions (*Pollock et al.* 2016).

Figure 1 exhibits an overview of MMS multiple crossings of the magnetosheath and the bow shock on Oct 25, 2015. Since the separation between the four MMS spacecraft (~ 15 km) is much shorter than the characteristic scales of the magnetosheath and bow shock, the MMS spacecraft see nearly the same large scale features of magnetosheath and bow shock structures. For this reason we show only data from MMS1 in Figure 1. Figure 1a-1f show that the magnetosheath is highly turbulent with large magnetic field fluctuations, low ion velocity, high ion density, and typical ion energies between ~ 100 eV and ~ several keV and electron energies between tens of eV and hundreds of eV. During the period from 08:40 UT to 09:40 UT, MMS crossed the bow shock four times (marked by blue dashed lines), and stayed in the magnetosheath in the time periods marked by red bars on the top of Figure 1 (the solar wind or foreshock periods are marked by black bars on the same figure). Figure 1g-1i zoom in a short period (40 seconds) of observations in the magnetosheath. It shows the presence of a magnetic hole with a magnetic depression marked by two black dashed lines.

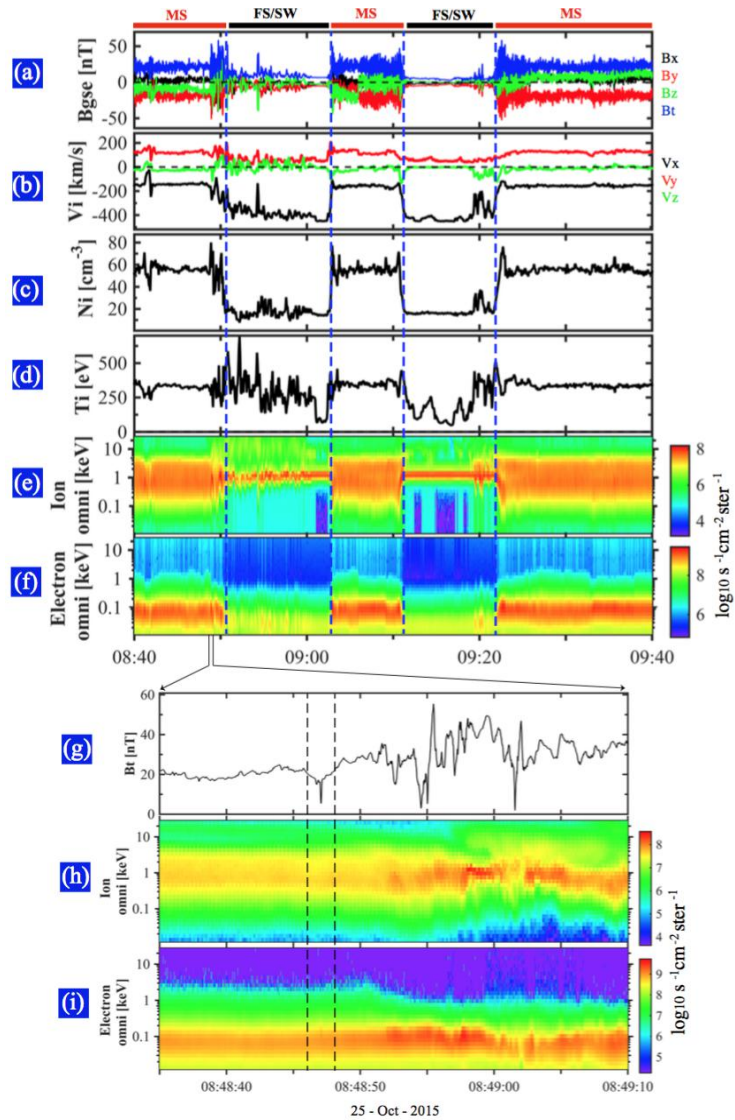

Figure 1. Overview observations from MMS1. (a) magnetic field, (b) ion velocity, (c) ion density, (d) ion temperature, (e) ion differential energy flux, and (f) electron differential energy flux. The crossings of bow shock are marked by blue dashed lines. The crossings of magnetosheath (MS) and foreshock/solar wind (FS/SW) are denoted by red bars and black bars on the top of (a), respectively. (g)-(i) present zoom-in observations of the magnetosheath, where (g) magnetic field, (h) ion differential energy flux, and (i) electron differential energy flux. The time interval of the observations of electron vortex magnetic hole is marked by two black dashed lines in (g)-(i).

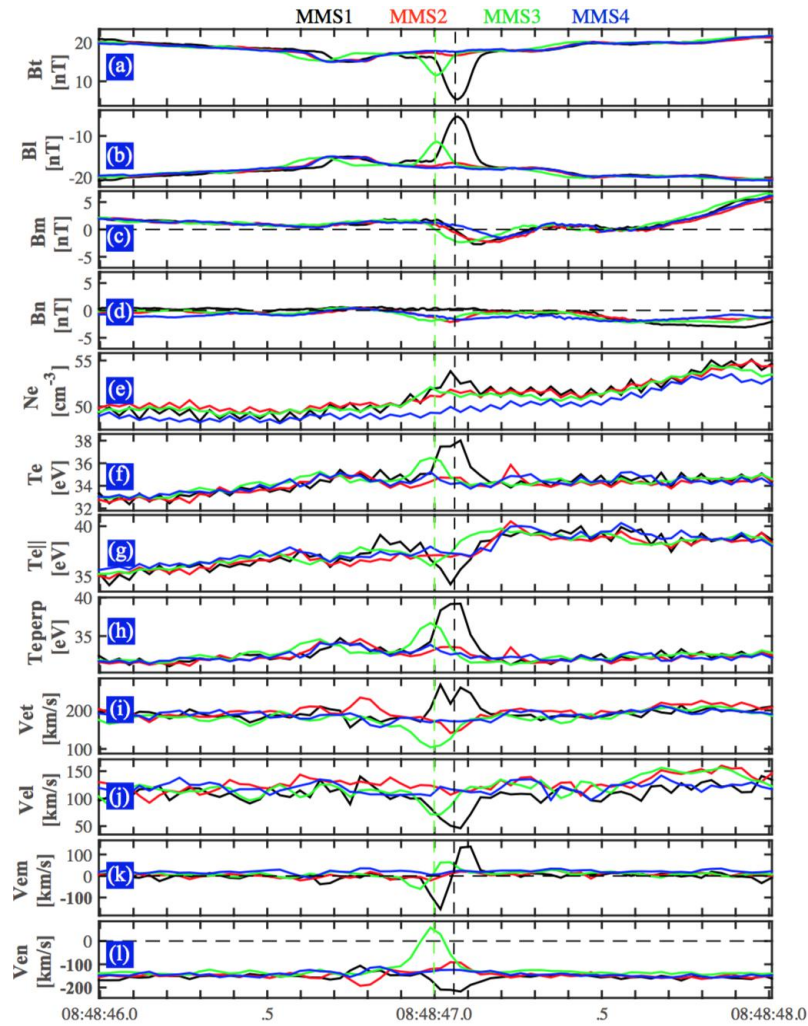

Figure 2. Detailed observations of the electron vortex magnetic hole in the magnetosheath turbulent plasma (vector data are given in LMN coordinates) from the four MMS spacecraft. (a)-(d) magnitude and three components of the magnetic field. (e) electron density, (f) electron temperature, (g) electron parallel temperature, (h) electron perpendicular temperature, and (i)-(l) magnitude and three components of the electron velocity. Black and green dashed lines mark the minimum magnetic field detected by MMS1 and MMS3 respectively.

Figure 2 gives detailed observations of the magnetic hole from 08:48:46 UT to 08:48:48 UT. All the data are presented in the LMN coordinates which were determined by minimum variance analysis (MVA) of the magnetic field from MMS1 (08:48:46.986 – 08:48:47.144 UT) [*Sonnerup and Scheible*, 1998]. In this coordinate system, **L** ([-0.33, 0.75, 0.58] GSE) is the maximum variation direction that can be considered as the axis of the magnetic hole, **N** ([-0.48, -0.66, 0.58] GSE) is the minimum variation direction, and **M** ([0.81, -0.09, 0.58] GSE) completes a right-handed orthogonal coordinate system. MMS3, MMS1 and MMS2 successively detected a reduction in the magnetic field strength (Figure 1a) that is essentially due to the field-aligned component (Figure 1b) around 08:48:47 UT. The magnetic depletion is up to 77% on MMS1, to which we generally refer as a magnetic hole. Figure 2d-2l displays the electron observations from the four spacecraft. The electron density (Figure 2e) and electron temperature increase (Figure 2f) inside the magnetic hole. However, the electron parallel temperature (with respect to the ambient magnetic field) decreases (Figure 2g), while the electron perpendicular temperature significantly increases (Figure 2h). MMS1 observed the strongest reduction in magnetic field along with the strongest decrease (resp. increase) of the electron parallel (resp. perpendicular) temperature. During the crossing of the magnetic hole, the electron velocities in $V_{em}$ component exhibited a bipolar variation, namely, $V_{em}$ changes from reverse to along **M**-direction, implying the possible existence of an electron vortex structure in the M-N plane inside the magnetic hole. By combining the different variations of the $V_{en}$ component on the different spacecraft (e.g., the increase of $V_{en}$ seen by MMS1, and the decrease of $V_{en}$ seen on MMS2 and MMS3), and the large ambient electron flow $V_{en}$, the magnetic hole can be interpreted as an electron vortex imbedded in the ambient plasma flow. In this scenario, one may infer that MMS1 is crossing one half of the electron vortex with positive $V_{en}$, while MMS2 and MM3 are crossing the other half with negative $V_{en}$. We will return to this point at the end of this Letter.

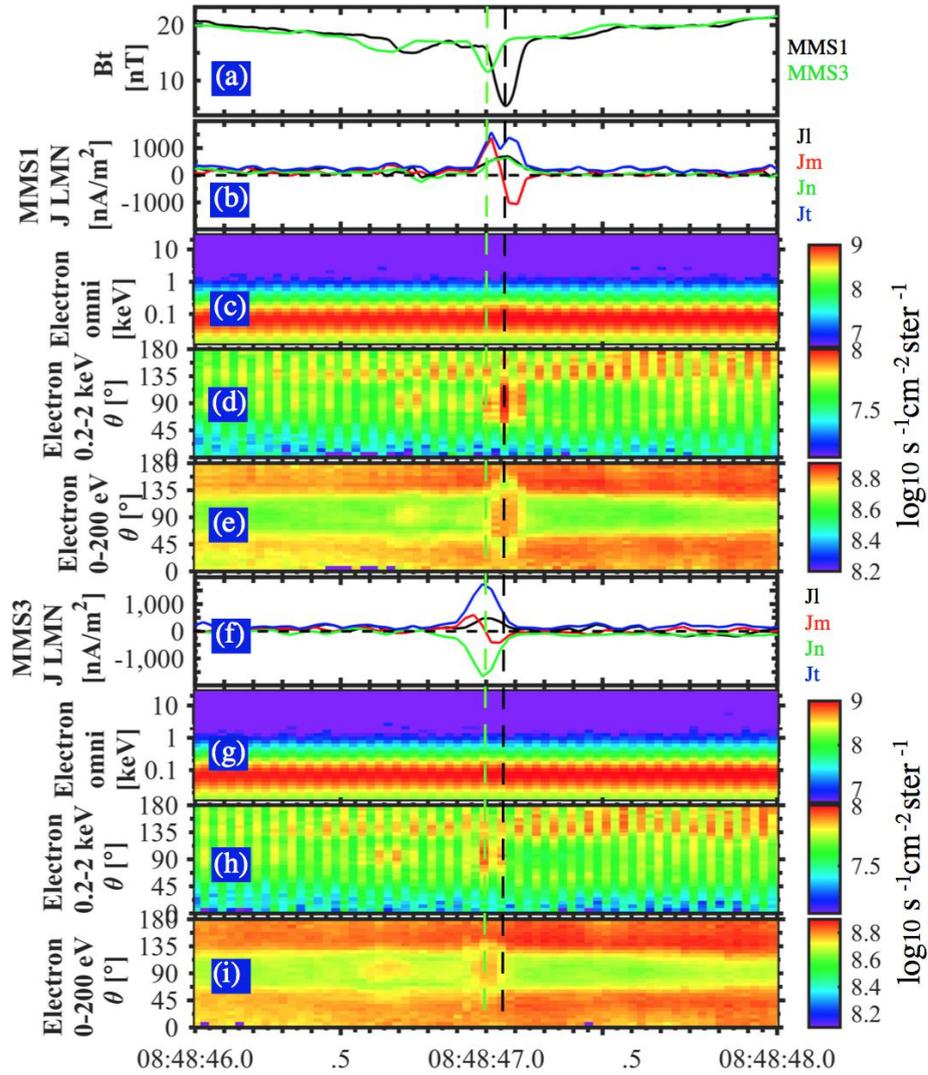

Figure 3. Current and particle observations of electron vortex magnetic hole. (a) magnitude of magnetic field from MMS1 and MMS3; (b) the current, (c) electron differential energy fluxes, (d) and (e) electron pitch angle distributions at energy band 0.2-2 keV and 0-200 eV respectively from MMS1; (f)-(i) is the same format as (c)-(e) but for MMS3. Black and green dashed lines mark the minimum magnetic field detected by MMS1 and MMS3 respectively.

Figure 3 shows the currents and electron pitch angle distributions associated with the identified magnetic hole from MMS1 and MMS3. The pitch angle distribution from 2 to 30 keV is not presented because the fluxes at these high energies are too low. The current is estimated at each spacecraft using the plasma measurements (i.e., $\boldsymbol{J} = ne(\boldsymbol{V}_i - \boldsymbol{V}_e)$), where $n$ is the plasma density, $e$ is the change, $V_i$ is the ion flow, and $V_e$ is the electron flow). All current components in Figures 3b and 3f are intense in the magnetic hole, and exhibit similar shapes opposite to the electron velocities in Figures 2i-2l, indicating that electrons carry most of the electric current. There is a dip in the total current in the center of the magnetic hole, and a peak in its outer region (Figure 3b). MMS3 only crossed the outer region of the magnetic hole, thus it detected only one peak of the current (Figure 3f). The current density profiles are well consistent with 2D and 3D PIC simulations (Figure 3 in Haynes et al. 2015 and Figure 8 in Roytershteyn et al. 2015).

The electron spectrogram does not show clear flux enhancement at high energies in the magnetic hole. Electrons behave differently inside and outside the magnetic hole. Bidirectional electrons are observed in the energy range 0-200 eV (Figure 3e and 3h) while anti-parallel electron flows are seen in the energy range 0.2-2 keV outside the magnetic hole (Figure 3d and 3g). The electrons have a similar behavior inside the magnetic hole, except the very clear enhancement in fluxes at ~ 90° pitch angle in the energy range 0-2 keV. Considering that no increase in fluxes is observed at high energies, one can deduce that these 90° pitch angle electrons are trapped and form an electron vortex within the magnetic hole. Note that very slight depressions in the magnetic field strength are detected before the magnetic hole (i.e., around 08:48:46.7 UT). These are also accompanied with flux enhancement around 90° pitch angles, suggesting that both MMS1 and MMS3 observe a similar vortex structure at that time.

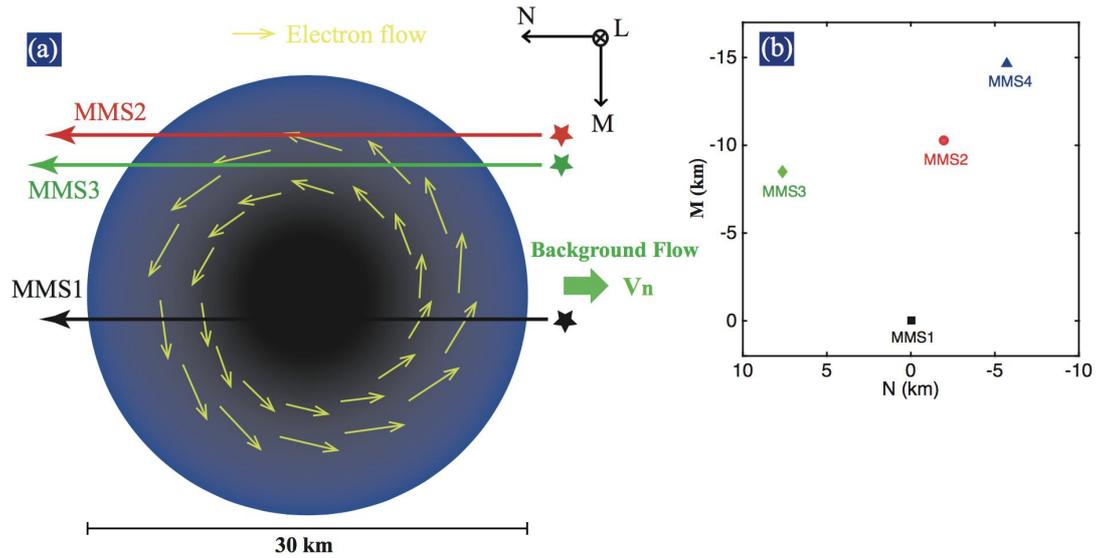

Figure 4. Cross-section of electron vortex magnetic hole and MMS position in M-N plane. (a) Schematic of electron vortex magnetic hole in M-N plane. The color code describes the amplitude of magnetic field (black: small; blue: large). Thick green arrow indicates the background flow $V_n$. Three thin arrows indicate the trajectories of MMS1 (black), MMS2 (red) and MMS3 (green). The black arrows show the electron flow in the cross-section. (b) The position of four MMS spacecraft in M-N plane.

Figure 4 provides a schematic of the observed magnetic hole by MMS. Figure 4a draws the nearly circular cross-section geometry of the magnetic hole, while Figure 4b presents the positions of the four spacecraft in the M-N plane (perpendicular to the axis of the magnetic hole). Given the average background velocities in Figures 2k and 2l, $V_{en} \sim -150$ km/s and $V_{em} \sim 0$ km/s, the magnetic hole was moving opposite to the **N**-direction. MMS3 was located in the most left in **N** direction, and MMS1 was in-between MMS3 and MMS2, ensuring the successive encounters of MMS3, MMS1, and MMS2 with the magnetic hole (their trajectories are marked by three horizontal colored arrows in Figure 4a). Moreover, the magnetic hole is expected to be in the form of an anticlockwise vortex, leading to variations of $V_{em}$ on MMS3, MMS1 and MMS2 from negative to positive direction, as well as a decrease in $|V_{en}|$ for MMS3 and MMS2 and an increase in $|V_{en}|$ for MMS1 (Figures 2k-2l). In order to estimate the scale of the magnetic hole in the M-N plane, one can assume that the magnetic hole moves with the ambient plasma flow (Sun et al., 2012). One can see that the electron velocity outside the magnetic hole is rather steady (Figure 2i-2l). Therefore, the width of the magnetic hole in the **N** direction $D_N$ is about 30 km using the equation $D_N = |V_{en}| \times dt$, where $dt \approx 0.2$ s is the duration of the crossing of the magnetic hole for MMS1. The estimated size $D_N$ is about 0.23 $\rho_i$ (~ 30 $\rho_e$), where $\rho_i$ and $\rho_e$ are the proton and electron gyroradius ($\rho_i \sim 127$ km and $\rho_e \sim 1$ km based on $|B| \sim 20$ nT, $n \sim 53$ cm$^{-3}$, $T_i \sim 310$ eV and $T_e \sim 33$ eV). The width in the **M**-direction can be estimated using the separation between the MMS spacecraft as shown in Figure 4b. MMS1 crosses the center of the magnetic hole, and MMS4 is located outside of the edge of the magnetic hole since it observed a slight decrease in $B_t$, yielding an approximate width of about 30 km in the **M** direction. This suggests that the electron-scale electron vortex magnetic hole has a circular cross-section in the M-N plane.

## 3. Discussions and conclusions

Small-scale magnetic holes with scales comparable to the proton gyroradius have been recently observed in the magnetospheric plasma sheet (e.g., Sun et al. 2012; Sundberg et al. 2015; Goodrich et al. 2016a, 2016b; Gershman et al. 2016). An electron anisotropy with an enhancement in the perpendicular energy fluxes at 90° pitch angles was observed in some magnetic holes (Sun et al. 2012; Sundberg et al. 2015; Gershman et al. 2016). In our work, using the high time resolution data from MMS, we also found an enhancement of energy flux at pitch angles 90° inside an electron-scale magnetic hole in the magnetosheath turbulent plasma. This implies that small-scale magnetic holes show similar electron behaviors in these different plasma environments. Goodrich et al (2016a) have reported that the Hall electron currents are responsible for the depression of magnetic field in the magnetic holes. Goodrich et al (2016b) mainly focused on the electric field of the magnetic holes, including the major magnetic hole and a secondary magnetic hole with slight depression ~20% and short duration ~0.5 s. Using the high time resolution particle data from MMS, Gershman et al (2016) have found that the currents carried by the electrons can be sufficient to account for depression of magnetic field in the magnetic hole. However, they observed the magnetic holes under the condition of electron temperature isotropy ($T_{e\perp} \sim T_{e\parallel}$) inside of the magnetic hole, which contrasts with the present observations (i.e., $T_{e\perp} > T_{e\parallel}$). As for the background plasma, the magnetic holes in the two studies were detected under similar electron plasma temperature conditions ($T_{e\parallel} > T_{e\perp}$), which may imply that both of them may be formed by the same mechanism.

Our comprehensive analysis sheds light on the physical mechanism that produces the observed magnetic hole. All the clues indicate the existence of an electron vortex magnetic hole which persists in a turbulent plasma where the ambient plasma has $T_{e\parallel} > T_{e\perp}$ (Haynes et al. 2015; Roytershteyn et al. 2015). The observed features, including density, electric current, electron temperature, electron vortex structure, trapped 90° pitch angle electrons and background plasma environment with $T_{e\parallel} > T_{e\perp}$, are all consistent with the predictions given by the simulations (Haynes et al. 2015; Roytershteyn et al. 2015).

Other generation mechanisms have been proposed for small-scale magnetic holes, including electron mirror mode or field-swelling instabilities (e.g., Gary and Karimabadi, 2006; Pokhotelov et al. 2013), tearing modes (Balikhin et al. 2012), and solitary waves (e.g., Baumgärtel 1999; Stasiewicz et al. 2003; Li et al. 2015; Yao et al. 2016). However, there are inconsistencies between the basic assumptions of the aforementioned mechanisms and the present observations. For example, in our event we have $T_i>T_e$ and $T_{e\|}/T_{e\perp}<1$ for the background plasma, which do not fulfill the theoretical threshold of the electron mirror instability ($T_{e\perp}/T_{e\|} >1+1/\beta_{e\perp}$) and field-swelling instabilities ($T_e>T_i$) (e.g., Gary and Karimabadi, 2006; Pokhotelov et al. 2013). The tearing mode may occur in localized regions of the cross magnetotail current sheet, and is thought to result in isotropic electrons in the magnetic hole (Balikhin et al. 2012), which is inconsistent with the background environment and anisotropic electron distributions found in the present study. Finally, the solitary wave theory is mainly appropriate in the weak nonlinear regime (Li et al. 2016; Yao et al. 2016), but may not be suitable for the strong nonlinear state of the magnetic hole in the magnetosheath turbulent plasma where magnetic field and plasma flow have large fluctuations.

In summary, we presented the detailed analysis of an electron vortex magnetic hole with a scale smaller than the proton gyroradius in the magnetosheath turbulent plasma using the MMS data. The magnetic hole has an electron-scale circular cross-section with a radius of 0.23 $\rho_i$. It is characterized by electron density and temperature increases, with an electron parallel temperature decrease but a perpendicular temperature increase ($T_{e\perp} > T_{e\parallel}$) inside the magnetic hole. The current density has a dip in the center of the magnetic hole and a peak at its outer edge. Electrons are trapped inside the magnetic hole, and form an electron vortex in the cross-section plane with an enhancement in the perpendicular electron flux at 90° pitch angles. All these observational features are well consistent with the simulation results of Haynes et al. (2015) and Roytershteyn et al. (2015). Our observations demonstrate that the coherent structures behaving as electron vortex magnetic holes can be generated in space plasma turbulence.

**Acknowledgement**

We thank the entire MMS team and instrument leads for data access and support. This work was supported by the National Natural Science Foundation of China (41374168, 41404132, 41574168, 41674161), Program for New Century Excellent Talents in University (NCET-13-0446), and China Postdoctoral Science Foundation Funded Project (2015T80830). SYH and FS acknowledge financial support from the project THESOW, grant ANR-11-JS56-0008 and from LABEX Plas@Par through a grant managed by the Agence Nationale de la Recherche (ANR), as part of the program "Investissements d'Avenir" under the reference ANR-11-IDEX-0004-02. Data is publicly available from the MMS Science Data Center at http://lasp.colorado.edu/mms/sdc/. Work at IRAP was supported by CNES and CNRS.